\documentclass{article}[12pt]
\usepackage[colorlinks,citecolor=red,urlcolor=blue,bookmarks=false,hypertexnames=true]{hyperref} 

\usepackage{cite}
\usepackage{bm}

\usepackage{diagbox}

\usepackage{chngcntr}

\usepackage{amssymb,amsmath,amsfonts}
\usepackage{float}
\usepackage{relsize}
\usepackage{graphicx}
\usepackage{tikz}
\usetikzlibrary{patterns}
\usetikzlibrary{shapes, arrows, shadows}

\tikzstyle{vertex} = [circle, draw, fill=blue!20, scale=1,auto=left]
\tikzstyle{vert} = [circle, draw, fill=blue!20, scale=.8,auto=left]
\tikzstyle{line} = [draw]

\usepackage{upgreek}

\usepackage{pgfplots}

\pgfplotsset{compat=newest}

\pgfdeclarepatternformonly[\LineSpace]{my north east lines}{\pgfqpoint{-1pt}{-1pt}}{\pgfqpoint{\LineSpace}{\LineSpace}}{\pgfqpoint{\LineSpace}{\LineSpace}}%
{
    \pgfsetlinewidth{0.4pt}
    \pgfpathmoveto{\pgfqpoint{0pt}{0pt}}
    \pgfpathlineto{\pgfqpoint{\LineSpace + 0.1pt}{\LineSpace + 0.1pt}}
    \pgfusepath{stroke}
}

\newdimen\LineSpace
\tikzset{
    line space/.code={\LineSpace=#1},
    line space=3pt
}

\def\be{\begin{equation}}
\def\ee{\end{equation}}
\def\bea{\begin{eqnarray}}
\def\eea{\end{eqnarray}}

\begin{document}

\begin{titlepage}
\date{\today}       \hfill

\begin{center}

\vskip .2in
{\LARGE \bf  OPE  in a generally covariant form}\\
\vspace{5mm}

\today
 
\vskip .250in

\vskip .3in
{\large Anatoly Konechny}

\vskip 0.5cm
{\it Department of Mathematics,  Heriot-Watt University\\
Edinburgh EH14 4AS, United Kingdom\\[10pt]
and \\[10pt]
Maxwell Institute for Mathematical Sciences\\
Edinburgh, United Kingdom\\[10pt]
}
E-mail: A.Konechny@hw.ac.uk
\end{center}

\vskip .5in
\begin{abstract}
We discuss the general covariance of operator product expansion in $D$-dimensional Euclidean 
conformal field theories. We propose to organise the expansion in  powers of geodesic distance between two insertion points and 
to use the tangent vector to the geodesic for contractions with tensor operators. For conformally flat manifolds we show by explicit calculation that 
certain curvature terms arise in the OPE. For example for $D>2$ the leading term of this type in the identity channel of OPE of two scalar primaries is proportional to the 
Schouten tensor. We further argue that the  terms we found are present for a general metric and are thus universal but there may be curvature terms at higher order in the expansion whose  coefficients are not determined by the flat space OPE.
The curvature  terms we discuss are of practical interest in conformal perturbation theory calculations on curved spaces.

\end{abstract}

\end{titlepage}

\renewcommand{\thepage}{\arabic{page}}
\setcounter{page}{1}
\large


\section{Introduction }
\renewcommand{\theequation}{\arabic{section}.\arabic{equation}}
\setcounter{equation}{0}
\large 

Operator product expansion (OPE)  is one of the most fundamental concepts in conformal field theory (see e.g. \cite{Hollands_Wald,Hollands,Frob,OPE_conv} for a general discussion). 
When considered on a flat Euclidean space ${\mathbb R}^{D}$ we can use the global conformal symmetry to argue that an  OPE 
of two scalar primary fields must take the form 
\be \label{flat_OPE}
{\cal O}_{i}(x_{1}) {\cal O}_{j}(x_2) = \sum_{k} C_{ij}^{(k)} \frac{1}{|x_1-x_{2}|^{\Delta_{i} + \Delta_{j} - \Delta_{k}}} 
P^{\nu_{1} \dots \nu_{r}}(\hat x_{12}^{\mu}) {\cal O}_{\nu_{1} \dots \nu_{r}}^{(k)}(x_1) \, .
\ee
Here the summation runs over all scaling fields (primary and descendants) ${\cal O}_{\nu_{1} \dots \nu_{r}}^{(k)}$ that are rank $r\ge 0$ tensors. In the OPE they are contracted with the $c$-number tensors $P^{\nu_{1} \dots \nu_{r}}(\hat x_{12}^{\mu})$ that are homogeneous functions of  coordinates of degree zero and are constructed from the unit length  vector 
\be
\hat x_{12}^{\mu} = \frac{x_{1}^{\mu} - x_{2}^{\mu}}{|x_1-x_2|} \, .
\ee

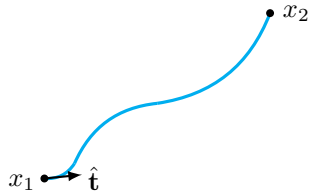
\begin{figure}[h!]
    \centering
   \begin{tikzpicture}[>=latex]
   
   \begin{scope}
     \draw[cyan, very thick] (0,0)  to[bend right] (0.4,0.2);
   \draw[cyan, very thick] (0.4,0.2)  to[bend left] (1.5,1);
    \draw[cyan, very thick] (1.5,1)  to[bend right] (3,2.2);
    \draw[black,thick,->] (0,0) --(0.5,0.05);
   \end{scope}
   \fill[black] (0,0)circle (1.5pt);
    \fill[black] (3,2.2)circle (1.5pt);
    \draw (-0.3,-0.05) node{$x_1$};
     \draw (3.35,2.2) node{$x_2$};
     \draw (0.66,0) node{$\hat {\bf t}$};
   
   \end{tikzpicture}
   \caption{Geodesic connecting two insertion points with $\hat {\bf t}$ -- the unit length tangent vector at $x_1$.}
   \label{geodesic_pic}
   \end{figure}

 In this paper we discuss how the form of  OPE changes when  our CFT lives on a surface with a general metric $g_{\mu \nu}$ and in general local  coordinates. We propose to organise the OPE in powers of the geodesic distance $\hat d=\hat d(x_1,x_2)$ between   the two insertion points and to replace the flat space tensors 
 $P^{\nu_{1} \dots \nu_{r}}(\hat x_{12}^{\mu})$ by covariant tensors constructed using the metric, curvature tensor and the tangent vector 
$\hat t^{\mu}$   along the geodesic taken at the expansion point, see Figure \ref{geodesic_pic} for illustration. 
We can write such an OPE as 
\be \label{curved_OPE}
{\cal O}_{i}(x_{1}) {\cal O}_{j}(x_2) = \sum_{k}\sum_{N=0}^{\infty} \hat C_{ij}^{(k,N)} \frac{1}{\hat d^{\Delta_{i} + \Delta_{j} - \Delta_{k}-N}} 
\hat P^{\nu_{1} \dots \nu_{r}}_{N}(\hat t^{\mu}, \hat g_{\alpha \beta}) {\cal O}_{\nu_{1} \dots \nu_{r}}^{(k)}(x_1) 
\ee
where the $c$-tensor $\hat P^{\nu_{1} \dots \nu_{r}}_{N}(\hat t^{\mu}, \hat g_{\alpha \beta})$ has dimension $N$ under locally constant Weyl rescalings (that is Weyl transformations for which the rescaling factor  is constant in a neighbourhood of $x=x_1$).
The advantage of this organisation is the general covariance. For concrete manifolds one can look for  a convenient choice of coordinates to do computations, e.g. one could work with Riemann normal coordinates. 

For conformally flat manifolds we can extract the OPE on a manifold with metric 
\be \label{Weyl_metric}
\hat g_{\mu \nu}  = \Lambda^{2}(x) \delta_{\mu \nu} 
\ee
from the flat space OPE (\ref{flat_OPE}) assuming our CFT is invariant under local Weyl transformations. 
The latter assumption means that the correlation functions of scalar primary fields ${\cal O}_{i}$ transform as 
\be  \label{Weyl}
\langle {\cal O}_{1}(x_1) \dots {\cal O}_{n}(x_n)\rangle_{\Lambda^2 \delta_{\mu \nu}} = 
\prod_{i=1}^{n} \Lambda^{-\Delta_{i}}(x_i) \langle {\cal O}_{1}(x_1) \dots {\cal O}_{n}(x_n)\rangle_{ \delta_{\mu \nu}}
\ee
where $\langle \dots \rangle_{g_{\mu \nu}}$ stands for a correlator on ${\mathbb R}^{D}$ with metric $g_{\mu \nu}$ and 
$\Delta_{i}$ are the scaling dimensions. As the 2- and 3-point functions are known for the standard Euclidean metric 
we can use (\ref{Weyl}) to find them for the metric (\ref{Weyl_metric}) and then extract the OPE (\ref{curved_OPE}).

One motivation for studying the covariant OPE is to get a description of singularities in the correlation functions on a curved space.
The latter is important for developing the conformal perturbation theory where the theory is often put on a cylinder that is curved for $D\ge 3$. 
To look at a concrete example let us consider the 3-dimensional cylinder $S_{R}^{2}\times {\mathbb R}$ with metric 
\be
\hat g_{\rm Cyl} = d\tau^2 + R^2 ds^2_{S^{2}} 
\ee
where $ds^2_{S^{2}}$ is the round sphere metric and $R$ is the radius of the sphere. 
This space is Weyl equivalent to the Euclidean space ${\mathbb R}^3$ with a puncture at the origin. 
We have 
\be \label{cyl_Weyl}
\hat g_{\rm Cyl}  =  \Lambda^2 ds^2_{\mathbb R^3}  \, , \quad \Lambda = R e^{-\tau/R} 
\ee
where $\tau = R\ln |x|$ relates $\tau$ to the Euclidean space norm.
Using (\ref{Weyl}) we obtain for the 2-point function of a primary field 
\be \label{2pt_cyl}
\langle {\cal O}(x_1){\cal O}(x_2)\rangle_{\rm Cyl} = 
\frac{1}{R^{2\Delta}(e^{(\tau_1-\tau_2)/R} + e^{-(\tau_1-\tau_2)/R} -2\cos \theta_{12})^{\Delta}}
\ee
where $\theta_{12}$ is the angle between the vectors $x_1$ and $x_2$ in Euclidean space. 
Putting $x_1$ at the north pole of our sphere and using standard spherical polar coordinates we 
have   $\theta_{12}=\theta_{1}-\theta_{2}$ -- the difference of polar angles. 
To generate a short distance expansion of (\ref{2pt_cyl}) 
it is convenient to introduce a rescaled coordinate $\sigma = R \theta$ and a complex variable 
\be
\xi = \frac{1}{R}[\tau_2-\tau_1 + i(\sigma_{2}-\sigma_{1})]\, , 
\ee
then we can rewrite the cylinder 2-point function as 
\be
\langle {\cal O}(x_1){\cal O}(x_2)\rangle_{\rm Cyl} =  \frac{1}{(R|\xi|)^{2\Delta} }F_{\Delta}(\xi) F_{\Delta}(\bar \xi)
\ee
where 
\be
F_{\Delta}(\xi) = \left(\frac{\xi}{e^{\xi}-1}\right)^{\Delta} e^{\frac{\Delta}{2}\xi} \, .
\ee
The latter function can be expanded in power series 
\be
F_{\Delta}(\xi) =  \sum_{n=0}^{\infty} B_{n}^{(\Delta)}(\Delta/2) \frac{\xi^{n}}{n!}
\ee
where $B_{n}^{(\alpha)}(\beta)$ denotes the generalised Bernoulli polynomials.
The leading correction is 
\bea \label{cyl_lead}
\langle {\cal O}(x_1){\cal O}(x_2)\rangle_{\rm Cyl} &= & 
 \frac{1}{(R|\xi|)^{2\Delta} }(1 -\frac{\Delta}{24}(\xi^2 + \bar \xi^2) + \dots  \nonumber \\
 &=& \frac{1}{( \sigma_{12}^2 + \tau_{12}^2)^{\Delta}}
 ( 1 + \frac{\Delta}{12 R^2} ( \sigma_{12}^2-\tau_{12}^2) + {\cal O}(1/R^4) )
\eea
Since the quantity $ \hat d^2 =\sigma_{12}^2 + \tau_{12}^2$ is exactly the square of the distance in the metric $\hat g_{\rm Cyl}$ ,
the leading term here can be interpreted as coming from the flat space OPE term proportional to the identity 
operator 
\be
{\cal O}(x_1){\cal O}(x_2) = \frac{1}{\hat d^{2\Delta}} + \dots 
\ee
The question we would like to address is -- what is the nature of the subleading terms in this expansion? Note that, although we chose $D=3$ for definiteness, in appropriate coordinates the above expansion holds on the cylinder for any $D$, including $D=2$ in which case the cylinder is flat. The nature of the subleading terms in this case is well known -- they come from the terms in the OPE that belong to the identity conformal tower. Due to the conformal anomaly those operators have non-vanishing 1-point functions on the cylinder. In three dimensions there is no conformal anomaly but the cylinder has curvature  so that it is natural to expect that 
the subleading terms come from some curvature terms (times the identity operator) that arise in the OPE. While in four dimensions there is a conformal anomaly the leading correction in (\ref{cyl_lead}) has dimension 2 and thus cannot be related to the conformal anomaly. 
By rewriting the OPE in a covariant form we will show that for $D>2$ the leading correction in the identity channel has the form 
\be \label{2pt_gen_correction_1}
 \ {\cal O}_{i}(x_1) {\cal O}_{j}(x_2) =   
 \frac{\delta_{ij}}{\hat d^{2\Delta_{i}}}\left( 1 + 
\frac{\Delta_{i}}{6}\hat d^2 \hat P_{\mu \nu}(x_1)\hat t^{\mu}\hat t^{\nu} + \dots \right)
\ee
where $\hat P_{\mu \nu}(x_1)$ stands for the value of the normalised Schouten tensor of the $\hat g$ metric:
\be \label{Schouten_1}
 \hat P_{\mu \nu} = \frac{1}{d-2}\left( \hat R_{\mu \nu} - \frac{1}{2(d-1)}\hat g_{\mu \nu} \hat R \right)\, . 
\ee
In particular, this universal term is responsible\footnote{While the leading correction term in the $1/R$ expansion in (\ref{cyl_lead}) is not exactly the same as the contribution of the Schouten tensor both terms coincide in the limit $|x_{12}|\to 0$.} for the leading correction to the cylinder 2-point function (\ref{cyl_lead}).

To illustrate the utility of  (\ref{2pt_gen_correction_1}) in conformal perturbation theory let us consider the leading  contribution to the free energy of the theory perturbed by ${\cal O}$. It is obtained by integrating the 2-point function (\ref{2pt_cyl}) over the infinite cylinder that gives \cite{Conf_pert1}
\be
\int \langle {\cal O}(x_1){\cal O}(x_2)\rangle_{\rm Cyl} \Lambda^{D} d^{D} x = \mathsmaller{R^{D-2\Delta} \pi^{\frac{D+1}{2}}2^{1-\Delta}
\frac{\Gamma(D/2-\Delta)\Gamma(\Delta/2)}{\Gamma^{2}(D/2-\Delta/2)\Gamma(1/2+\Delta/2)}} \, .
\ee
The pole at $\Delta = D/2$ signals the logarithmic divergence in position space that is associated with the leading singularity in the OPE  coming from the identity field. The pole at $\Delta = D/2 + 1$ is another logarithmic divergence that receives two contributions: one  from the subleading divergence 
 associated with the identity operator obtained by expanding the volume measure factor $\Lambda^{D}$ and the second one comes from 
the subleading term in the OPE expansion that according to  (\ref{2pt_gen_correction_1}) comes from the Schouten tensor.  
With the cylinder being a homogeneous manifold we can cancel such a divergence with the identity field counterterm. On non-homogeneous manifolds the situation may be different requiring curvature counterterms that can be identified using the covariant OPE expansion. 

In section 2 of the paper we explain how to obtain the leading curvature corrections in the OPE. Formula (\ref{cov_exp2}) is the main result of this paper.  In section 3 we discuss the $D=2$ case 
in more detail. In section 4 we offer some concluding remarks. The appendix contains more details on the derivative expansion for the geodesics.



\section{The leading curvature terms in covariant OPE}

Our goal  is to rewrite  the OPE of two  primary operators  on a conformally flat space with metric (\ref{Weyl_metric}) in the covariant form (\ref{curved_OPE}). As in the flat space the OPE coefficients can be obtained by applying the OPE inside a 3-point function. 
In this paper we focus on the scalar primary fields and their descendants arising in the OPE of two scalar primaries. 
We assume that our scalar primaries ${\cal O}_{i}$ are normalised so that the 2-point functions in the flat Euclidean space are
\be
\langle {\cal O}_{i}(x_1) {\cal O}_{j}(x_2)\rangle_{\delta_{\mu\nu}} =   \frac{\delta_{ij}}{|\Delta x|^{2\Delta_{i}}} \equiv   \frac{\delta_{ij}}{d^{2\Delta_{i}}} \, . 
\ee
The 3-point functions then have the standard form 
\be
\langle {\cal O}_{i}(x_1) {\cal O}_{j}(x_2){\cal O}_{k}(x_3)\rangle_{\delta_{\mu\nu}} =
 \frac{C_{ijk}}{d_{12}^{\Delta_{i}+\Delta_{j}-\Delta_{k}}d_{13}^{\Delta_{i}+\Delta_{k}-\Delta_{j}}d_{23}^{\Delta_{j}+\Delta_{k}-\Delta_{i}}}
\ee
where $d_{ij}$ are the Euclidean distances between $x_{i}$ and $x_{j}$, and $C_{ijk}$ are the structure constants which are totally symmetric in the indices. Using (\ref{Weyl}) we obtain from the above expressions the 2- and 3-point functions on the space with conformally flat metric (\ref{Weyl_metric}).
Instead of using the 
3-point functions at finite points it is convenient to put one insertion at infinity:
\bea \label{3pt_inf}
\langle k|  {\cal O}_{j}(x_2){\cal O}_{i}(x_1) \rangle_{\Lambda^2\delta_{\mu \nu}} &\equiv&  \lim_{x_{3}\to \infty} d_{13}^{2\Delta_{k}} 
 \langle {\cal O}_{k}(x_3) {\cal O}_{j}(x_2) {\cal O}_{i}(x_1)\rangle_{\Lambda^2\delta_{\mu \nu}}
 \nonumber \\
 &=& \frac{C_{ijk}\Lambda^{-\Delta_{i}}(x_1)\Lambda^{-\Delta_{j}}(x_2)}{d^{\Delta_{i}+\Delta_{j} - \Delta_{k}}}
\eea
where $d\equiv d_{12}$.
To extract the covariant OPE expansion we expand (\ref{3pt_inf}) in derivatives of the Weyl rescaling function $\Lambda(x)$ with respect to
the conformal coordinates $x^{i}$ specified by  (\ref{Weyl_metric}). The derivatives in their turn are expressed  in terms of the covariant quantities such as the geodesic distance $\hat d$ computed in the metric $\hat g_{\mu \nu}$ and the tangent vector to the 
corresponding geodesic $\hat t^{\mu}$. We assume that the insertion points $x_1$  and $x_2$ are sufficiently close to each other so that there is a unique geodesic connecting them. The geodesic equation can then be solved iteratively order by order in  derivatives of $\Lambda$. We will use the following expansions (see Appendix A for  details of the derivation)
\be \label{d-expansion2}
\frac{1}{d} = \frac{\Lambda_{1}}{\hat d}\left( 1 + \frac{\hat d}{2}(\eta \cdot \hat t\, ) 
+ \hat d^2\left[ \frac{\hat m}{6}
-\frac{7}{24} (\eta \cdot \hat t\, )^2 + \frac{5}{24} \hat \eta^2 \right] \right) + {\cal O}(\partial^3) \, , 
\ee
\be \label{Lambda-expansion2}
\Lambda^{-1}(x_2) = \Lambda_{1}^{-1}\left(1-\hat d(\eta \cdot \hat t) + 
\frac{\hat d^2}{2}[ 3 (\eta \cdot \hat t)^2 - \hat \eta^2 -\hat m]  \right) + {\cal O}(\partial^3) \, ,
\ee
\be
\hat t^{\mu} = \frac{1}{d\Lambda_1}\left(\Delta x^{\mu}[1+\frac{1}{2}(\eta\cdot \Delta x)] - \frac{d^2}{2}\eta^{\mu}\right)   + {\cal O}(\partial^2)\, . 
\ee
where ${\cal O}(\partial^{n})$ stands for terms containing  $n$ and higher number of  derivatives of $\Lambda$. 
The above expansions contain derivatives of $\Lambda$ evaluated at $x_1$ for which we use the following notation 
\be
 \eta_{\mu} = \partial_{\mu} \ln \Lambda(x_1) \, , \qquad   \hat \eta^2 = \hat g^{\mu \nu}(x_1)\eta_\mu\eta_\nu = \Lambda_{1}^{-2} \delta^{\mu\nu}\eta_\mu\eta_\nu\, , 
\ee
\be
m_{\mu \nu}=\partial_{\mu}\partial_{\nu}\ln \Lambda(x_1) \, , \qquad \hat m = m_{\mu \nu}\hat t^{\mu} \hat t^{\nu} \, . 
\ee
We have also abbreviated 
\be
\Lambda_{1}=\Lambda(x_1)\, , \qquad \Delta x^{\mu} = x_{2}^{\mu}-x_{1}^{\mu} \, .
\ee

Using (\ref{d-expansion2}) 
(\ref{Lambda-expansion2}) and retaining terms up to two derivatives we obtain an expansion of (\ref{3pt_inf}) in powers 
of $\hat d$. The result is the following expansion 
\bea \label{3pt_inf_exp2}
&& \langle k|  {\cal O}_{j}(x_2){\cal O}_{i}(x_1) \rangle_{\Lambda^2\delta_{\mu \nu}}  =
 \frac{C_{ijk}\Lambda^{-\Delta_{i}}(x_1)\Lambda^{-\Delta_{j}}(x_2)}{d^{\Delta_{i}+\Delta_{j} - \Delta_{k}}} \nonumber \\
 && = \frac{C_{ijk}\Lambda_{1}^{-\Delta_{k}}}{\hat d^{\Delta_{i}+\Delta_{j}-\Delta_{k}}}
 \Bigl[ 1 + \mathsmaller{\frac{\Delta_{i}-\Delta_{j}-\Delta_{k}}{2}}\hat d(\eta\cdot \hat t) + \mathsmaller{\frac{\Delta_{i}-\Delta_{k}-2\Delta_{j}}{6}}\hat d^2 \hat m \nonumber \\
 && + \mathsmaller{\frac{5(\Delta_{i}-\Delta_{k})-7\Delta_{j}}{24}}\hat d^2 \hat \eta^2 + \left[ 
 \mathsmaller{\frac{(\Delta_{i}+\Delta_{j}-\Delta_{k})^2}{8}}-\mathsmaller{\frac{(6\Delta_j+5)(\Delta_{i}+\Delta_{j}-\Delta_{k})}{12}}
 + \mathsmaller{\frac{\Delta_{j}(\Delta_{j}+2)}{2}}\right] \hat d^2 (\eta\cdot \hat t)^2 + \dots \Bigr] \nonumber \\
 && 
\eea
valid to second order in derivatives.
We  would like to match this formula to a covariant OPE expansion performed inside the correlator (\ref{3pt_inf}). 
We expect part of the expansion to come from the known flat space OPE that to the second order in derivatives is
\bea
&&{\cal O}_{j}(x_2) {\cal O}_{i}(x_1) = \sum_{k}\frac{C_{ijk}}{|\Delta x|^{\Delta_{i} + \Delta_{j}-\Delta_{k}}}\Bigl(  
{\cal O}_{k}(x_1) + \alpha_{ijk} \Delta x^{\mu}\partial_{\mu}{\cal O}_{k}(x_1)  \nonumber \\
&& + \beta_{ijk}\Delta x^{\alpha}\Delta x^{\beta}\partial_{\alpha}\partial_{\beta}{\cal O}_{k}(x_1)    + \gamma_{ijk} \Delta x^2 \delta^{\alpha \beta} \partial_{\alpha}\partial_{\beta}{\cal O}_{k}(x_1) +\dots \Bigr)
\eea
where
\be
\alpha_{ijk} = \mathsmaller{\frac{\Delta_{j}+ \Delta_{k} - \Delta_{i}}{2\Delta_{k}}} \, , \qquad \beta_{ijk} = 
\mathsmaller{\frac{(\Delta_{j} + \Delta_{k} - \Delta_{i})(\Delta_{j} + \Delta_{k} - \Delta_{i}+2)}{8\Delta_{k}(\Delta_{k} + 1)}} \, ,
\ee
\be
\gamma_{ijk} = \mathsmaller{\frac{(\Delta_{j} + \Delta_{k} - \Delta_{i})(\Delta_{i} + \Delta_{k} - \Delta_{j})}{8\Delta_{k}(\Delta_{k} + 1)(D-2\Delta_{k} -2)} }\, .
\ee
Covariantising the above terms and adding two curvature terms with coefficients $A_{ijk}$, $B_{ijk}$
we expect to reproduce   (\ref{3pt_inf_exp2}) from the OPE  
\bea \label{cov_exp2}
&&{\cal O}_{j}(x_2) {\cal O}_{i}(x_1) = \sum_{k}\frac{C_{ijk}}{\hat d^{\Delta_{i} + \Delta_{j}-\Delta_{k}}}\Bigl(  
{\cal O}_{k}(x_1) + \alpha_{ijk} \hat d \hat t^{\mu}\partial_{\mu}{\cal O}_{k}(x_1)  \nonumber \\
&& + \beta_{ijk}\hat d^2 \hat t^{\alpha}\hat t^{\beta}\nabla_{\alpha}\partial_{\beta}{\cal O}_{k}(x_1)    + 
\gamma_{ijk} \hat d^2  \hat g^{\alpha \beta} \nabla_{\alpha}\partial_{\beta}{\cal O}_{k}(x_1) \nonumber \\
&& + A_{ijk}\hat d^2 \hat R {\cal O}_{k}(x_1) + B_{ijk}\hat d^2 \hat R_{\alpha \beta}\hat t^{\alpha}\hat t^{\beta}  {\cal O}_{k}(x_1) +  \dots \Bigr)
\eea 
where $\nabla_{\alpha}$ denote the covariant derivatives for the metric $\hat g$. The Ricci tensor and the scalar curvature for the metric $\hat g$ are denoted as $\hat R_{\alpha \beta}$ and $\hat R$ respectively.
To compute the contributions of the OPE terms to  the correlation function (\ref{3pt_inf}) we use 
\be \label{1pt1}
\langle k|{\cal O}_{r}(x_1)\rangle = \delta_{kr} \Lambda_{1}^{-\Delta_{k}}\, , 
\ee
\be \label{1pt2}
\langle k|\hat t^{\mu}\partial_{\mu} {\cal O}_{r}(x_1)\rangle = -\delta_{kr}\Lambda_{1}^{-\Delta_{k}}\Delta_{k}(\eta\cdot \hat t)\, , 
\ee
\be \label{1pt3}
\langle k| \hat t^{\alpha}\hat t^{\beta}\nabla_{\alpha}\partial_{\beta}{\cal O}_{r}(x_1)\rangle = 
\delta_{kr}\Lambda^{-\Delta_{k}}_{1}\left[-\Delta_{k}\hat m + \Delta_{k}(\Delta_{k} + 2)(\eta\cdot\hat t)^2 -\Delta_{k}\hat \eta^2\right]\, , 
\ee
\be \label{1pt4}
\langle k| \hat g^{\alpha \beta} \nabla_{\alpha}\partial_{\beta}{\cal O}_{r}(x_1)\rangle = 
\delta_{kr}\Lambda^{-\Delta_{k}}_{1}\left[-\Delta_{k}\hat M + \Delta_{k}(\Delta_{k}+2-D)\hat \eta^2 \right]
\ee
where 
\be
\hat M = \hat g^{\alpha \beta} m_{\alpha \beta} = \Lambda_{1}^{-2}\delta^{\alpha \beta}\partial_{\alpha} \partial_{\beta}\ln \Lambda_{1} \, .
\ee
Substituting (\ref{cov_exp2}) into (\ref{3pt_inf}) and using formulae (\ref{1pt1}) - (\ref{1pt4}) we obtain an expansion that we set equal to 
 (\ref{3pt_inf_exp2}).   Matching the coefficients at $\hat \eta^2$, $(\eta\cdot \hat t)^2$, $\hat m$, and $\hat M$ we obtain four linear equations for the two coefficients 
$A_{ijk}$, $B_{ijk}$. For $d>2$ the unique solution is 
\be \label{A}
A_{ijk} =\mathsmaller{\frac{1}{24(D-1)(D-2)}\left[  \mathsmaller{-3}\frac{(\Delta_{j}+\Delta_{k}-\Delta_{i})(\Delta_{i} + \Delta_{k}-\Delta_{j})(D-2-\Delta_{k})}
{(\Delta_{k}+1)(D-2-2\Delta_{k})} \mathsmaller{+ \Delta_{k}-\Delta_{i} - \Delta_{j} }\right]} \, ,
\ee
\be \label{B}
B_{ijk} = \mathsmaller{\frac{(\Delta_{j}+\Delta_{k}-\Delta_{i})(\Delta_{i} + \Delta_{k}-\Delta_{j})}{8(D-2)(\Delta_{k}+1)}
+\frac{\Delta_{i}+\Delta_{j}-\Delta_{k}}{12(D-2)}} \, .
\ee
We note that these coefficients are manifestly symmetric under the interchange of $i$ with $j$ as expected for OPE coefficients of scalar fields. 
Formula (\ref{cov_exp2}) with the coefficients  (\ref{A}), (\ref{B}) is our main result. While we used conformal flatness to calculate the coefficients 
 $A_{ijk} $, $B_{ijk}$ they must be universal with the corresponding terms present for a general metric.
This is due to the fact that  on one hand for conformally flat manifolds a general metric formula must reduce to (\ref{cov_exp2}) but on the other hand the obstruction tensors such as the Cotton tensor in $D=3$ and the Weyl tensor in $D=4$ cannot appear at this order in the derivative expansion due to their natural scaling dimensions.  

Specialising to the identity channel: $\Delta_{k}=0, \Delta_{i}=\Delta_{j}$, $\alpha_{ijk}=\beta_{ijk}=\gamma_{ijk}=0$ 
 we obtain from the above the contribution 
 \bea \label{2pt_gen_correction}
 {\cal O}_{i}(x_1) {\cal O}_{j}(x_2) &=&   
 \frac{\delta_{ij}}{\hat d^{2\Delta_{i}}}\left( 1 + 
\frac{\Delta_{i}}{6}\hat d^2( (\eta \cdot \hat t)^2 - \frac{\hat \eta^2}{2}-\hat m) + \dots \right)
\nonumber \\
& =&\frac{\delta_{ij}}{\hat d^{2\Delta_{i}}}\left( 1 + 
\frac{\Delta_{i}}{6}\hat d^2 \hat P_{\mu \nu}(x_1)\hat t^{\mu}\hat t^{\nu} + \dots \right)
\eea
where $\hat P_{\mu \nu}(x_1)$ stands for the value of the normalised Schouten tensor of the $\hat g$ metric:
\bea \label{Schouten}
&& \hat P_{\mu \nu} = \frac{1}{d-2}\left( \hat R_{\mu \nu} - \frac{1}{2(d-1)}\hat g_{\mu \nu} \hat R \right)\nonumber \\
&& =
 \partial_{\mu}\ln\Lambda \partial_{\nu} \ln \Lambda -\frac{\delta_{\mu\nu}}{2}\delta^{\alpha \beta}\partial_{\alpha} \ln \Lambda \partial_{\beta} \ln \Lambda  - \partial_{\mu} \partial_{\nu} \ln \Lambda \, .
\eea

It is instructive to take a look at the case of formula (\ref{2pt_gen_correction}) for the cylinder (\ref{cyl_Weyl}). We obtain 
\be \label{Phat_cyl}
 \hat P_{\mu \nu}(x_1)\hat t^{\mu}\hat t^{\nu} = \frac{1}{R^2}\left( \frac{|x_2|^2}{|\Delta x|^2}\sin^2(\theta_{12}) - \frac{1}{2}\right) 
\ee
where the vector lengths are the Euclidean ones. Using the cylinder coordinates $\tau$ and $\sigma$ introduced in 
the introduction section it is straightforward to check that  
\be
\hat d^2  \hat P_{\mu \nu}(x_1)\hat t^{\mu}\hat t^{\nu} = \frac{1}{2}(\sigma_{12}^2 - \tau_{12}^2 + {\cal O}(\hat d^4))
\ee
so that in the $R\to \infty $ limit with finite coordinates $\theta, \tau$ the leading correction in (\ref{cyl_lead}) is reproduced by the general expression (\ref{2pt_gen_correction}). However, at finite $\Delta x^{\mu} $ the $1/R$ expansion given in  (\ref{cyl_lead}) is not the same as the covariant expansion in 
(\ref{2pt_gen_correction}). The exact expression in (\ref{Phat_cyl}) contains infinitely many corrections in $\tau_{12}$ and in $\theta_{12}$ in 
comparison to the leading term in  (\ref{cyl_lead}). They come from the dependence of $\hat t^{\mu}$ on $x_{2}$.

\section{The $D=2$ case }
The covariant OPE coefficients (\ref{A}), (\ref{B}) we found in the previous section are singular at $D=2$.  
For $D=2$ all fields in the identity conformal tower receive anomalous contributions under Weyl transformation 
and thus typically have non-trivial 1-point functions on curved backgrounds. Similarly, there are contributions to the 
3-point functions (\ref{3pt_inf}) coming from Virasoro descendants such as, at the leading order,   the normal ordered products 
$(T{\cal O}_{k})\equiv L_{-2}{\cal O}_{k}$ 
and  $(\bar T{\cal O}_{k})\equiv \bar L_{-2}{\cal O}_{k}$.
These anomalous terms
absorb some of the derivatives of $\Lambda$  present in (\ref{3pt_inf_exp2}). The remaining derivatives can be put into terms proportional to the scalar curvature. 
While the 1-point functions are standard and have been extensively discussed in the literature, to the best of our knowledge the curvature term has  not been discussed before.

It is convenient to use the flat complex coordinates $z, \bar z$ in which 
$\hat g = \Lambda^2 dzd\bar z$. 
Taking into account the Virasoro descendants we replace the covariant OPE (\ref{cov_exp2})  by 
\bea \label{cov_OPE_D=2}
&& {\cal O}_{i}(z_1, \bar z_1){\cal O}_{j}(z_2, \bar z_2) = \sum_{k}\frac{C_{ijk}}{\hat d^{\Delta_{i} + \Delta_{j}-\Delta_{k}}}\Bigl(  
{\cal O}_{k}(z_1, \bar z_1) \nonumber \\
&& + \mathsmaller{\frac{h_{j}+h_{k}-h_{i}}{2h_{k}} } \hat d (\hat t^{z}\partial_{z}{\cal O}_{k}(z_1, \bar z_1)  + \hat t^{\bar z}\partial_{\bar z}{\cal O}_{k}(z_1, \bar z_1)) + \beta_{1} \hat d^2 [(\hat t^{z})^2{\cal O}_{k}^{(2)}(z_1, \bar z_1)  + (\hat t^{\bar z})^2\bar {\cal O}_{k}^{(2)}(z_1, \bar z_1)   ] \nonumber \\
&& + 
\beta_{2}  \hat d^2  \hat t^{\alpha} \hat t^{\beta}  \nabla_{\alpha}\partial_{\beta}{\cal O}_{k}(z_1, \bar z_1) + 
\beta_{3}  \hat d^2 \hat g^{\alpha \beta}  \nabla_{\alpha}\partial_{\beta}{\cal O}_{k}(z_1, \bar z_1) 
+ A_{ijk}^{\mathsmaller{\rm D=2}} \hat d^2 \hat R {\cal O}_{k}(z_1, \bar z)  +  \dots \Bigr) 
\eea 
where 
\be
{\cal O}_{k}^{(2)} = L_{-2}{\cal O}_{k} - \frac{3}{2(2h_{k}+1)}L_{-1}^2{\cal O}_{k} \, , \qquad \bar {\cal O}_{k}^{(2)} = \bar L_{-2}{\cal O}_{k} - \frac{3}{2(2h_{k}+1)}\bar L_{-1}^2{\cal O}_{k}
\ee
are the quasiprimary level 2 fields  and the derivative coefficients are those of the flat space OPE
\bea
&& \beta_{1}  =   \frac{-3(h_{j}+h_{k}-h_{i})(h_{i}+h_{k}-h_{j}+1) + 2(2h_{k}+1)(2h_{j}+h_{k}-h_{i})}{c(2h_{k}+1) + h_{k}(16h_{k}-10)} \, , \nonumber \\
&& \beta_{2}  =  \frac{(h_{j}+h_{k}-h_{i})(h_{j}+h_{k}-h_{i}+1)}{4h_{k}(2h_{k}+1) }\, , \nonumber \\
&& \beta_{3}= -\frac{(h_{i}+h_{k}-h_{j})(h_{j}+h_{k}-h_{i})}{16h_{k}^2 (2h_{k}+1)} \, .
\eea
These coefficients can be found using standard methods, see e.g.   \cite{yellow_book}.
For the ease of comparison with  2D CFT literature we use the conformal weights $h_{k} = \Delta_{k}/2$, and $c$ is the central charge. 
The contributions of the descendant fields ${\cal O}_{k}^{(2)}$ and $\bar {\cal O}_{k}^{(2)}$ in (\ref{cov_OPE_D=2}) may not appear fully covariant, however they are as they
correspond to a contraction of two copies of  $\hat t^{\mu}$ with a traceless tensor.

Note that  we have 
\be \label{deltazz}
\hat d \hat t^{z} = \Delta z + {\cal O}(\partial) \, , \quad \hat d \hat t^{\bar z} = \Delta \bar z + {\cal O}(\partial) \, . 
\ee
 Since in (\ref{cov_OPE_D=2}) we are expanding to second order in derivatives the ${\cal O}(\partial)$ terms in  (\ref{deltazz}) only contribute at the first order derivatives in (\ref{cov_OPE_D=2}). 

To find the coefficient $A_{ijk}^{\mathsmaller{\rm D=2}}$ at the curvature term (and to check the entire ansatz) we substitute (\ref{cov_OPE_D=2}) into the 3-point function (\ref{3pt_inf}) and compare to (\ref{3pt_inf_exp2}).  To evaluate the contributions of the OPE terms 
 we use formulae (\ref{1pt1}) - (\ref{1pt4})  along with 
\be \label{quasiprimary}
\langle k| {\cal O}^{(2)}_{k}(z_1, \bar z_1)\rangle_{\Lambda^2\delta_{\mu \nu}} = -\Lambda_{1}^{-2h_{k}} 
\frac{c(2h_{k}+1)+h_{k}(16h_{k}-10)}{6(2h_{k}+1)}(\partial_{z}^2\ln \Lambda - (\partial_{z}\ln \Lambda)^2) 
\ee
and the analogous formula for $ \bar {\cal O}^{(2)}_{k}$
that are obtained\footnote{While the formula derived by Gaberdiel in \cite{Gaberdiel} is for a local diffeomorphism it is equivalent to a local Weyl transformation of the metric in view of conformality.} using the general transformation formula of \cite{Gaberdiel}. Performing the calculation we find again an overdetermined system of linear equations that has a unique solution 
\be
A_{ijk}^{\mathsmaller{\rm D=2}} = \frac{h_{i} + h_{j}}{24} + \frac{h_{k}^{2} - 3(h_{i}-h_{j})^2}{48h_{k}} \, .
\ee

In the identity channel we have $h_{k}=0$ and $h_{i} = h_{j}$. Formula (\ref{quasiprimary}) reduces to the well-known 1-point 
functions of the stress-energy tensor   
\be
\langle T\rangle_{\Lambda^{2}\delta_{\mu \nu}} = -2\pi  \langle T_{z z}\rangle_{\Lambda^{2}\delta_{\mu \nu}} = \frac{c}{6} [(\partial_{z}\ln \Lambda)^2 - \partial_{z}^2 \ln \Lambda] \, , 
\ee
\be
\langle \bar T\rangle_{\Lambda^{2}\delta_{\mu \nu}} = -2\pi  \langle T_{\bar z \bar z}\rangle_{\Lambda^{2}\delta_{\mu \nu}} = \frac{c}{6} [(\partial_{\bar z}\ln \Lambda)^2 - \partial_{\bar z}^2 \ln \Lambda] 
\ee
 that follow from the conformal anomaly. The corresponding terms in the OPE are
 \bea
&   {\cal O}_{i}(x_1) {\cal O}_{j}(x_2) = 
  \frac{\delta_{ij}}{\hat d^{2\Delta_{i}}}\Bigl[ 1 -\frac{2\pi \Delta}{c} \hat d^2 [(\hat t^{zz})^2T_{zz}(x_1) + (\hat t^{\bar z\bar z})^2 T_{\bar z\bar z}(x_1)]
 \nonumber \\
 &    + \frac{\Delta}{24} \hat d^2 \hat R(x_1) + \dots  \Bigr]
\eea
 where the ellipsis stands for the non-identity primary channel contributions  and for terms that are higher order in derivatives. 
 
 \section{Concluding remarks}
 Here we would like to point out how our result given in formula (\ref{cov_exp2}) could be generalised. Firstly, for conformally flat metrics, one could calculate
 the higher order terms in (\ref{cov_exp2})  that may be particularly interesting to do in $D=4$ where the conformal anomaly is expected to contribute at the 4-th order in derivatives. Moving on to primary fields with spin one can calculate the leading order curvature corrections standing at such terms in the OPE of two scalars as well as the leading corrections to the OPE of fields with non-trivial spin. 
 
 Another interesting direction is to consider the covariant OPE on non-conformally flat manifolds. The ambient space formalism has been adapted 
 in \cite{Skenderis1}, \cite{Skenderis2} to cover the form of 2- and 3-point functions on such spaces.  The advantage of the ambient space formalism is maintaining the 
 explicit covariance under Weyl transformations (see also \cite{kravchuk_etal} where the issue of Weyl covariance is tackled in the context of fusion of conformal defects).   Perhaps those 
 results may be used to extract a generalisation of formula (\ref{cov_exp2}).
  It would be particularly interesting to see if there are 
 terms in the OPE proportional to the Cotton tensor in D=3 or to the Weyl tensor in D=4 for which the OPE coefficients cannot be derived from the flat space OPE \cite{inprogress}. Such terms may be linked to the anomalous terms in the Weyl transformation  discussed in \cite{Luty_etal}. 
 
It would be also interesting to study the convergence of OPE on curved spaces. One obvious idea may be to use the results of  \cite{OPE_conv} 
extending them to conformally flat manifolds. This suggests using flat metric spheres as regions of convergence. It may be more desirable though, especially with the view of extending the results to general metrics,  to keep everything covariant and to use instead the geodesic length spheres for the curved metric. We leave these questions to future work.

\begin{center}
{\bf \large Acknowledgements} 
\end{center}
I would like to thank Matt Walters and Petar Tadi\'c for discussions and Slava Rychkov for comments on the draft of the paper.

\appendix
\renewcommand{\theequation}{\Alph{section}.\arabic{equation}}
\counterwithin{figure}{section}
\setcounter{equation}{0}

\section{Derivative expansion for geodesic distance  on conformally flat manifolds}

Let us consider a coordinate patch $U$ with coordinates $x^{\mu}$ of a $d$-dimensional conformally flat manifold. 
We consider two metrics on the patch -- the flat metric $\sum_{i} (dx^{\mu})^2$ and the conformally rescaled metric 
with the tensor
\be
\hat g_{\mu \nu} = \Lambda^{2}(x) \delta_{\mu \nu} \, .
\ee
Let $x_{1}, x_{2} \in U$ be two points with unique geodesic $x^{\mu}(t)$, $t\in [0,1]$  in metric $\hat g$ between them. 
We can generate an expansion of $x(t)$ in derivatives of the scale factor $\Lambda$:
\be \label{geodesic_exp}
x^{\mu}(t) = x_{0}^{\mu}(t) + \delta_{1}x^{\mu}(t) +  \delta_{2}x^{\mu}(t) + \dots 
\ee 
where 
\be
x_{0}^{\mu}(t) = x_{1}^{\mu} + \Delta x^{\mu} t \, , \qquad \Delta x^{\mu} = x_{2}^{\mu} - x_{1}^{\mu} 
\ee
is the flat metric geodesic and $ \delta_{n}x^{\mu}(t)$ contains order $n$ derivative terms. 

The geodesic equation is 
\be \label{geodesic_eq}
\frac{d^{2} x^{\mu}}{dt^2} = \dot x^2 \partial^{\mu} \Lambda - 2 \dot x^{\mu} (\dot x^{\nu} \partial_{\nu} \Lambda)  
\ee
where the dots stand for derivatives in $t$ and we raise or lower indices with the flat metric $\delta_{\mu \nu}$ 
so that $ \dot x^2 = \dot x^{\nu}\dot x_{\nu}$ and $\partial_{\mu} \Lambda = \partial^{\mu} \Lambda$. 
By substituting  the expansion (\ref{geodesic_exp}) into (\ref{geodesic_eq}), truncating in derivatives and solving 
order by order we find
\be \label{delta1}
\delta_{1}x^{\mu}(t) = (t -t^2)\xi^{\mu} \, , 
\ee  
\bea \label{delta2}
&& \delta_{2}x^{\mu}(t) = 2\left( \frac{t^3}{3}-\frac{t^2}{2} + \frac{t}{6} \right)((\eta \cdot \xi)\Delta x^{\mu} + (\eta\cdot \Delta x) \xi^{\mu} 
-(\xi\cdot \Delta x)\eta^{\mu})  \nonumber \\
&& + \frac{1}{3}(t-t^3)(m\Delta x^{\mu} - \frac{1}{2}\Delta x^2 m^{\mu}_{\nu}\Delta x^{\nu})  
\eea
where the inner products $a\cdot b$ are taken with the flat metric and the following constant (i.e. $t$-independent) tensors appear
\be
 \eta_{\mu} = \partial_{\mu} \ln \Lambda(x_1) \, , \qquad \xi^{\mu} = (\eta_{\nu}\Delta x^{\nu})\Delta x^{\mu} - \frac{1}{2}\Delta x^2 \eta^{\mu} \, , 
\ee
\be
m_{\mu \nu}=\partial_{\mu}\partial_{\nu}\ln \Lambda(x_1) \, , \qquad m=m_{\mu \nu}\Delta x^{\mu}\Delta x^{\nu} \, . 
\ee
The geodesic length between $x_1$ and $x_2$ is given by 
\be \label{ghat-length}
\hat d= \int\limits_{0}^{1}\! \!dt\, \Lambda(x^{\mu}(t))\sqrt{\dot x^{\mu}(t) \dot x_{\mu}(t) } \, .
\ee
To compute this to the second order in derivatives we use (\ref{delta1}), (\ref{delta2}) and the Taylor expansion of 
$\Lambda(x)$ centred at $x^{\mu}=x_{1}^{\mu}$. To the second order in derivatives we have 
\be \label{Lambda-Taylor}
 \Lambda(x^{\mu}(t)) = \Lambda_1 +\Lambda_1 \eta_{\mu} ( \Delta x^{\mu} t +  \delta_{1} x^{\mu}(t)) + \frac{\Lambda_{1}t^2}{2} \Delta x^{\mu} \Delta x^{\nu} 
 ( \eta_{\mu}\eta_{\nu} + m_{\mu \nu}) + {\cal O}(\partial^3)
\ee
where $\Lambda_{1}=\Lambda(x_1)$. Substituting (\ref{Lambda-Taylor}) into (\ref{ghat-length}) and integrating we obtain 
to second order in derivatives
\be
\hat d = d \Lambda_1\left( 1+ \frac{1}{2} (\eta \cdot \Delta x) + \frac{m}{6}  + \frac{5}{24} (\eta \cdot \Delta x)^2 - \frac{1}{24} (\delta x)^2 \eta^2 \right) + {\cal O}(\partial^3)
\ee
where $d= \sqrt{\Delta x^2}$ is the distance in the flat metric.  This formula can be inverted to obtain $d$ as an expansion in powers of $\hat d$. 
This gives the following formula which we  find is particularly useful in CFT applications 
\be \label{d-expansion}
\frac{1}{d} = \frac{\Lambda_{1}}{\hat d}\left( 1 + \frac{\hat d}{2}(\eta \cdot \hat t\, ) 
+ \hat d^2\left[ \frac{\hat m}{6}
-\frac{7}{24} (\eta \cdot \hat t\, )^2 + \frac{5}{24} \hat \eta^2 \right] \right) + {\cal O}(\partial^3)
\ee
where  
\be
\hat t^{\mu} = \frac{\dot x^{\mu}(0)}{\sqrt{\Lambda_{1}^{2} \dot x^{\nu}(0)\dot x_{\nu}(0)}} 
\ee
stands for the unit length tangent vector to the geodesic at $x_1$ 
and 
\be
\hat \eta^2 = \hat g^{\mu \nu}(x_1)\eta_\mu\eta_\nu = \Lambda_{1}^{-2} \delta^{\mu\nu}\eta_\mu\eta_\nu \, , \qquad 
\hat m = m_{\mu \nu}\hat t^{\mu} \hat t^{\nu} \, .
\ee
To first order in derivatives we have
\be
\hat t^{\mu} = \frac{1}{d\Lambda_1}\left(\Delta x^{\mu}[1+\frac{1}{2}(\eta\cdot \Delta x)] - \frac{d^2}{2}\eta^{\mu}\right)   + {\cal O}(\partial^2)\, . 
\ee
This formula can be inverted to obtain 
\be
\Delta x^{\mu}=\hat d \hat t^{\mu}(1-\hat d (\eta \cdot \hat t)) + \frac{\hat d^2}{2}\hat \eta^{\mu} + {\cal O}(\partial^2)\, , \qquad \hat \eta^{\mu} = \Lambda_{1}^{-2} \eta_{\mu} \, .
\ee
Using the last formula we obtain a derivative expansion for the inverse scale factor at the other end point of the geodesic 
\be \label{Lambda-expansion}
\Lambda^{-1}(x_2) = \Lambda_{1}^{-1}\left(1-\hat d(\eta \cdot \hat t) + 
\frac{\hat d^2}{2}[ 3 (\eta \cdot \hat t)^2 - \hat \eta^2 -\hat m]  \right) + {\cal O}(\partial^3) \, .
\ee



\end{document}